\documentclass{IEEEtran}
\usepackage{cite}
\usepackage{amsmath,amssymb,amsfonts}
\usepackage{graphicx}
\usepackage{textcomp,nicefrac}
\usepackage{upgreek}
\usepackage{color}
\usepackage{booktabs}

\def\BibTeX{{\rm B\kern-.05em{\sc i\kern-.025em b}\kern-.08em
T\kern-.1667em\lower.7ex\hbox{E}\kern-.125emX}}
\begin{document}
\onecolumn
\thispagestyle{empty}\setcounter{page}{0}%

{\LARGE \begin{flushleft} \textbf{IEEE Copyright Notice}\end{flushleft}}
\vspace{1.1cm}
{\large \begin{flushleft} \copyright $\,$2022 IEEE.  Personal use of this material is permitted. Permission from IEEE must be obtained for all other uses, in any current or future media, including reprinting/republishing this material for advertising or promotional purposes, creating new collective works, for resale or redistribution to servers or lists, or reuse of any copyrighted component of this work in other works \end{flushleft}}
\vspace{1.1cm}
{\LARGE \begin{flushleft} DOI: 10.1109/TNS.2022.3222544\end{flushleft}}

\newpage
\twocolumn

\bstctlcite{IEEEexample:BSTcontrol}
\title{Personal Dosimetry in Direct Pulsed Photon Fields with the Dosepix Detector}
\author{\textbf{Dennis Haag, Sebastian Schmidt, Patrick Hufschmidt, Gisela Anton, Rafael Ballabriga, Rolf Behrens, Michael Campbell, Franziska Eberle,  Christian Fuhg, Oliver Hupe, Xavier Llopart, J{\"u}rgen Roth, Lukas Tlustos, Winnie Wong, Hayo Zutz and Thilo Michel}
\thanks{Paper submitted for review on 29.04.2022. This project is funded by the Deutsche Forschungsgemeinschaft (DFG, German Research Foundation) - 394324524.}
\thanks{Dennis Haag, Sebastian Schmidt, Patrick Hufschmidt, Gisela Anton, Franziska Eberle, and Thilo Michel are with the Erlangen Centre for Astroparticle Physics, Friedrich-Alexander Universit{\"a}t Erlangen-N{\"u}rnberg, 91058 Erlangen, Germany. e-mail: dennis.den.haag@fau.de.}
\thanks{Rolf Behrens, Christian Fuhg, Oliver Hupe, J{\"u}rgen Roth, and Hayo Zutz are with the Physikalisch-Technische Bundesanstalt (PTB), 38116 Braunschweig, Germany.}
\thanks{Rafael Ballabriga, Michael Campbell, Xavier Llopart, and Lukas Tlustos are with CERN, 1211 Geneva, Switzerland.}
\thanks{Winnie Wong was with CERN, 1211 Geneva, Switzerland. She is now with Mercury Systems, 1212 Geneva, Switzerland.}
}

\maketitle

\begin{abstract}
First investigations regarding dosimetric properties of the hybrid, pixelated, photon-counting Dosepix detector in the direct beam of a pulsed photon field (RQR8) for the personal dose equivalent $H\mathrm{_p(10)}$ are presented. The influence quantities such as pulse duration and dose rate were varied, and their responses were compared to the legal limits provided in PTB-A 23.2. The variation of pulse duration at a nearly constant dose rate of about 3.7$\,$Sv/h shows a flat response around 1.0 from 3.6$\,$s down to 2$\,$ms. A response close to 1.0 is achieved for dose rates from 0.07$\,$Sv/h to 35$\,$Sv/h for both pixel sizes. Above this dose rate, the large pixels (220$\,\upmu$m edge length) are below the lower limit. The small pixels (55$\,\upmu$m edge length) stay within limits up to 704$\,$Sv/h. The count rate linearity is compared to previous results, confirming the saturating count rate for high dose rates.
\end{abstract}


\IEEEpeerreviewmaketitle

\section{Introduction}
\IEEEPARstart{D}{osimetry} in pulsed photon fields with active electronic personal dosemeters (APDs) is an important topic of the last decade. The advantages of APDs compared to their passive counterparts are direct readability of the dose and an active warning of the wearer if pre-defined dose/dose rate thresholds are surpassed. An active warning is, for example, an alarm sound, a flashing light, or vibration of the dosemeter. They can help increase the wearers' awareness of unintentional high exposures and allow them to react to the situation directly. A pulsed radiation field is defined as ionizing radiation with a constant dose rate for pulse durations shorter than 10$\,$s \cite{ISO/TS}. With this definition, most X-ray tubes in medical applications are classified as pulsed radiation emitters \cite{PField}. The staff of interventional radiology, interventional cardiology (IR/IC), or veterinary medicine is not exposed to the direct radiation field during X-ray examinations. They stand next to the patient, i.e., in the scattered radiation field where dose rates of 0.1 to 3$\,\%$ of the dose rate in the direct radiation field can be expected \cite{10.1093/rpd/ncq351}. Nonetheless, it is essential to investigate the performance of APDs for dose rates of several hundreds of Sv/h. These dose rates occur in the case of situations in which the wearer of the dosemeter accidentally gets exposed by the direct beam of a pulsed radiation field. For example, the relevant radiation field characteristics in IR/IC are a pulse duration of 1-20$\,$ms, a dose rate in the direct radiation field of 2-360$\,$Sv/h, and of 5$\,$mSv/h to 10$\,$Sv/h in the scattered radiation field \cite{Clair}. Latter dose rate is stated for a dosemeter worn above the lead apron of the operator. The energy range of the spectra in the scattered beam of the radiation field for interventional procedures is between 20-100$\,$keV for peak high voltages of up to \mbox{120$\,$kV \cite{Clair}}. In this work, Dosepix \cite{Wong} is utilized in a system consisting of three Dosepix detectors to investigate its dose rate and pulse duration dependence for the personal dose equivalent $H\mathrm{_p(10)}$ in the direct beam of a pulsed photon field. It is tested whether a dosemeter demonstrator consisting of three Dosepix detectors fulfills the requirements in the direct beam of a pulsed photon field, in particular, for high peak pulse dose rates and short radiation pulse durations. When speaking of dose rate and pulse duration in the following, the dose rate in the pulse and the pulse width of a single pulse are meant.

\section{State of the art APDs}
A survey regarding the use of active personal dosemeters in hospitals showed that the predominantly used detector type are silicon diodes in IR/IC and Geiger-Muller tubes in nuclear medicine \cite{Ciraj_Bjelac_2018}. Two of the most commonly used silicon diode APDs were tested in an RQR8 reference radiation field \cite{Hupe}. It was found that they perform well for dose rates of a few Sv/h but become increasingly insensitive for higher dose rates. Those types of dosemeters need correction factors for dead-time loss. These correction factors become inefficient for pulse durations below 100$\,$ms \cite{Hupe}. Further tests in pulsed photon fields were performed in \cite{10.1093/rpd/ncz173} investigating dosemeters from the questionnaire in \cite{Ciraj_Bjelac_2018}. The tests showed that only 3 of 10 APDs fulfilled the requirement of a dose rate above 1$\,$Sv/h within the margin of $\pm$20$\%$ for the dose indication relative to the reference dose for continuous radiation. The maximum achieved dose rate within limits was 4$\,$Sv/h in an RQR8 reference photon field. Another study compares APDs to their passive counterparts in pulsed photon fields and \mbox{hospitals \cite{10.1093/rpd/ncz253}}. APDs (silicon diodes) show an under-response due to pulsed radiation compared to passive dosemeters, which are not influenced by pulsed radiation. The tested dosemeters perform well in the scattered radiation field in the hospital but show a response below 20$\,\%$ of the reference dose in the direct beam of the radiation field (dose rate of 74$\,$Sv/h in the pulse). The same dosemeters show a response below 70$\,\%$ relative to a continuous field in the direct beam of an RQR8 reference field at 5$\,$Sv/h and 10$\,$Sv/h. One of the three dosemeters shows a 50$\,\%$ response at already 1$\,$Sv/h. All in all, the main issues of APDs in pulsed photon fields are high peak pulse dose rates, as they occur in the direct beam of the radiation field, and short radiation pulse durations, where dead-time correction factors become inefficient. Worst case scenario is that the detector is exposed to high dose rates in the direct beam of a pulsed radiation field but does not respond at all. It gives the wearer a false sense of security, particularly that no noticeable radiation is present. A comprehensive overview of radiation dosemeter types is given in \cite{Ravotti}. Alternative active personal dosemeter systems in IR consist of CMOS image sensors as presented in \cite{CMOS_conti, Omar_2017}. In this work, the detector type of interest are hybrid photon-counting pixel detectors. Their principle of dosimetry is elaborated in \cite{Dosis}.

\section{The Dosepix detector}
\begin{figure}[!t]
\centering
\includegraphics[width=3.6in]{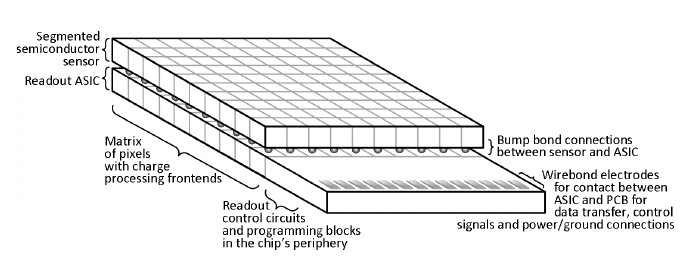}
\caption{Schematic depiction of the basic structure of the Dosepix detector. The image is taken from \cite{WinniePhD}.}
\label{fig:pm}
\end{figure}

Dosepix is a hybrid, pixelated, photon-counting X-ray detector. The hybrid design consists of an application-specific integrated circuit (ASIC) and a semiconductor sensor layer pixel-wisely connected via bump bonds to the ASIC. A schematic depiction of the basic structure is shown in \mbox{Figure \ref{fig:pm}}. A fully depleted 300$\,\upmu$m thick silicon sensor is used with an applied bias voltage of 100$\,$V. The pixel layout comprises 16$\times$16 square pixels with 220$\,\upmu$m pixel pitch with a p-in-n doping profile. The upper two and lower two rows of the sensor pixel matrix consist of small pixels with an edge length of 55$\,\upmu$m, while the remaining 12 rows consist of larger pixels with an edge length of 220$\,\upmu$m. The smaller pixels detect fewer events than the larger pixels and therefore have a lower tendency for pile-up, which allows applications at high-flux conditions. The Dosepix can be operated in 3 different programmable modes: the photon-counting mode, the integration mode, and the energy-binning mode. Here, the Dosepix detector is used in the latter one, which is specifically designed for dosimetry applications. The energy bin edges are individually programmed for each pixel. The Dosepix operates dead-time-free using the rolling-shutter principle. A single column is read out at a time while the rest of the matrix continues to process signals. This is a significant advantage in practical applications in pulsed photon fields, where radiation pulses are random in time. A single column consists of 12 large and 4 small pixels. The energy-binning mode pixelwise counts events in one of 16 histogram bins according to the deposited energy of the event. The total number of registered events is determined by summing the number of events per bin for each pixel type individually, resulting in two total energy histograms. Each energy histogram is multiplied by a dose equivalent and pixel size specific set of conversion factors to yield the personal dose equivalent $H_{\mathrm{DPX}}^{x}$. The dose for a single detector is calculated via \cite{Dosis}

\begin{equation}
	H_{\mathrm{DPX}}^\mathrm{x} = \sum\nolimits_{i=1}^{16} k_\mathrm{i}^\mathrm{x} N_\mathrm{i}^\mathrm{x}
\end{equation}

\noindent where $N_\mathrm{i}^\mathrm{x}$ denotes the number of events in energy bin $i$ for the pixel type $x$, and $k_\mathrm{i}^\mathrm{x}$ its corresponding conversion factor. The total dose is the sum of the dose contributions of the three detectors. Two separate dose values are calculated, one for the small and one for the large pixels. The statistical uncertainty of $H_{\mathrm{DPX}}^{x}$ is calculated following \cite{GUM} via a propagation of uncertainty assuming the uncertainty of $N_\mathrm{i}^\mathrm{x}$ as $u(N_\mathrm{i}^\mathrm{x}) = \sqrt{N_\mathrm{i}^\mathrm{x}}$:

\begin{equation}
	u(H_{\mathrm{DPX}}^{x}) = \sqrt{ \sum\nolimits_{i=1}^{\mathrm{16}} (k_\mathrm{i}^\mathrm{x} \sqrt{N_\mathrm{i}^\mathrm{x}})^2 }.
\end{equation}

\noindent Information regarding the characterizations of Dosepix with X-rays and analog test-pulses can be found in \cite{Ritter}, and measurements of the count rate linearity in dependence of the dose rate can be found in \cite{Zang}. A dosimetry system consisting of 3 Dosepix detectors is utilized as described in \cite{Hp10}, where the energy and angular dependence in continuous photon fields were already presented. Each of the three detectors is covered by a filter cap of a different shape and material. The filter caps are an aluminum half-sphere of 2$\,$mm thickness, a tin half-sphere of 1$\,$mm thickness and an aluminum cylinder with a thin aluminum foil of 0.25$\,$mm thickness on its top which has a hole above the sensor. The filter caps are optimized for the angular dependence of the personal dose equivalent $H\mathrm{_p(10)}$.

\section{Methods}
The tests in pulsed photon fields were performed in collaboration with and at the PTB (Physikalisch-Technische Bundesanstalt Braunschweig) using its X-ray unit for pulsed radiation GESA (GEpulste Strahlungs Anlage) presented in \cite{KlammerGesa}. The chosen reference radiation field is the medical radiation quality RQR8 with a tube voltage of 100$\,$kV filtered with 3.36$\,$mm aluminum and a mean energy (fluence) of 51$\,$keV \cite{IEC-61267}. Each measurement was repeated 2 or 3 times for statistical purposes. According to International Organization for Standardization (ISO) standards \cite{ISO-4037-3}, the dosemeter was irradiated on the center front of an ISO water slab phantom. The phantom is necessary because the operational quantity for individual monitoring of strongly penetrating radiation, $H\mathrm{_p(10)}$, is measured \cite{dos23}. It is defined at a depth of 10$\,$mm in the body, and is used for whole body dosemeters. The phantom serves as substitute of the human body in the radiation field and considers its backscattering. However, to achieve very high dose rates up to 1080$\,$Sv/h, the dosemeter had to be irradiated relatively close to the X-ray tube resulting in small field diameters down to 8.5$\,$cm. In these cases the dosimetry system was irradiated without the ISO water slab phantom which has a 30$\,$cm$\times$30$\,$cm cross-section and a thickness of 15$\,$cm. Correction factors were determined for measurements with and without the phantom and measurements with a small (8.5$\,$cm) and a large (42.0$\,$cm) field diameter. The correction factors are stated in Table \ref{T1} and were used to correct all measurements to the equivalent of the dosimetry system being placed on the ISO water slab phantom that is completely irradiated to guarantee $H\mathrm{_p(10)}$ conditions. The quantity of interest is the change of the response $R$ relative to the response at reference conditions $R_0$, i.e., the normalized response. The response is defined by the ratio of the indication (calculated dose with Dosepix) and the reference dose determined by monitor ionization chambers which are practically independent of the dose rate and pulse duration. The change of the normalized response has to fulfill the following condition according to \cite{PTB}

\begin{equation}
   1+f_{\mathrm{min}} \leq  R_{\mathrm{Norm}}^{i} = \frac{R_i}{R_{\mathrm{0}}} = \frac{H_{\mathrm{DPX}}^{i} H_{\mathrm{ref}}^{\mathrm{0}}}{H_{\mathrm{ref}}^{i} H_{\mathrm{DPX}}^{\mathrm{0}}} \leq  1+f_{\mathrm{max}}
\end{equation}

\noindent with $R_i$ being the response at measurement $i$, $H_{\mathrm{DPX}}$ the dose measured by the Dosepix dosimetry system, $H_{\mathrm{ref}}$ the reference dose, and $f_{\mathrm{min}}$ and $f_{\mathrm{max}}$ depending on the influence quantity (see Table \ref{TableAnf}). The statistical uncertainty of the normalized response is calculated via:

\begin{equation}
    u(R_{\mathrm{Norm}}^{i}) = R_{\mathrm{Norm}}^{i}
    \sqrt{ \left( \frac{u(H_{\mathrm{DPX}}^{i})}{H_{\mathrm{DPX}}^{i}} \right)^2 +
    \left( \frac{u(H_{\mathrm{DPX}}^{0})}{H_{\mathrm{DPX}}^{0}} \right)^2 }.
\end{equation}

\noindent The minimum requirements for the dose rate and pulse duration for conformity assessment are shown in Table \ref{TableAnf}. The limits of the statistical uncertainty of the measured dose for $H\mathrm{_p(10)}$ dosemeters regarding X-radiation are between 5-15$\,\%$ depending on the applied reference dose \cite{PTBc}. Test conditions with respect to pulsed radiation are stated in \cite{IEC-63050:2019}.

\begin{table}[!b]
\caption{Correction factors for the phantom influence and the field-diameter influence}
\label{T1}
\centering
\begin{tabular}{c|c|c}
\hline
 & & \\
 Pixel & Presence of Phantom & Field diameter\\
 size [$\,\upmu$m] & correction & correction\\
\hline
 & & \\
 55 & 1.148$\pm$0.004& 1.033$\pm$0.007 \\
 220 & 1.168$\pm$0.002& 1.024$\pm$0.003\\
 & & \\
\hline
\end{tabular}
\end{table}

\begin{table}[!b]
\caption{Minimum requirements for conformity assessment according to PTB-A 23.2 \cite{PTB} for $H\mathrm{_p(10)}$ dosemeters}
\label{TableAnf}
\centering
\begin{tabular}{c|c|c|c}
\hline
 & & &\\
Quantity & Minimum rated & Reference & $f_{\mathrm{min}}$, \\
& range of use&value&$f_{\mathrm{max}}$\\
\hline
 & & &\\
 & & &\\
 Dose rate&0.1$\,\upmu$Sv/h & 1$\,$mSv/h & -0.13, 0.18\\
 & to 1$\,$Sv/h & &  \\
 & & &\\
\hline
 & & &\\
Radiation pulse & & Response at & \\
duration & 1$\,$ms to 10$\,$s & continuous  & -0.2, 0.2 \\
 & & radiation &\\
 & & &\\
\hline
\end{tabular}
\end{table}

\section{Results and Discussion}

\subsection{Dependence on the pulse duration}
\begin{figure}[!t]
\centering
\includegraphics[width=3.5in]{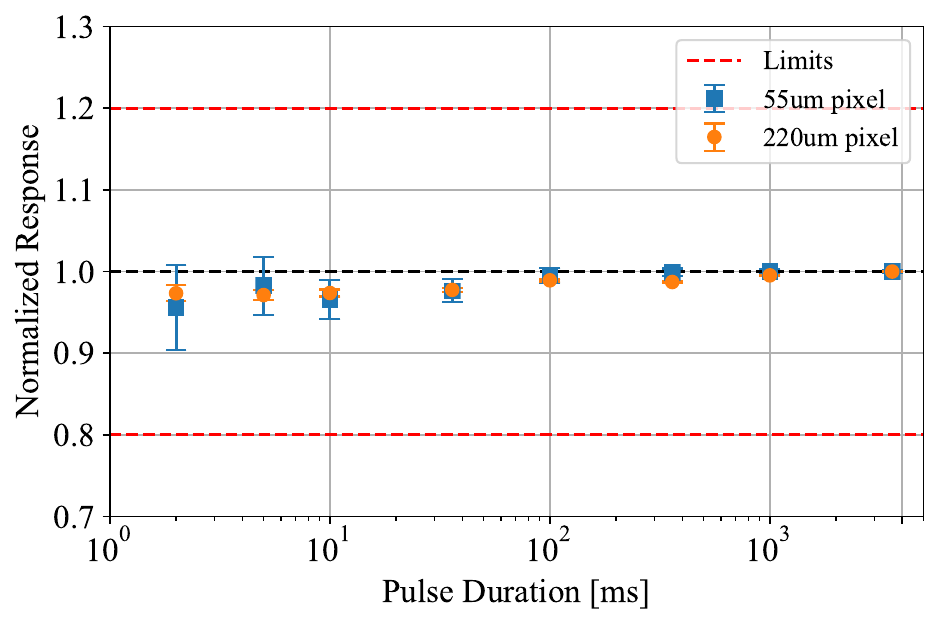}
\caption{Normalized response for $H\mathrm{_p(10)}$ in an RQR8 photon reference field. Influence quantity is the pulse duration at nearly constant dose rate. The uncertainty bars are calculated via (4).}
\label{pD}
\end{figure}

The pulse duration was varied between 2$\,$ms and 3.6$\,$s, while the dose rate was held nearly constant with an average of about 3.7$\,$Sv/h. The usual pulse duration is within the interval of several milliseconds in medical X-ray facilities. Due to the dead-time-free measurement of the Dosepix, no dependency on pulse duration is expected. Figure \ref{pD} shows the normalized response for both pixel sizes. The response value at 3.6$\,$s was chosen as reference point $R_\mathrm{0}$. The normalized response is flat within the margin of the uncertainty. The uncertainty increases for lower pulse durations due to fewer registered events in the energy bins. The number of registered events decreases because the reference dose is reduced with the pulse duration to hold the dose rate nearly constant for all irradiations. The reference dose values range from about 2.3$\,\upmu$Sv at 2$\,$ms to about 3.5$\,$mSv at 3.6$\,$s. For 2.3$\,\upmu$Sv an uncertainty limit of 15$\%$ is tolerated \cite{PTBc}. Here, the large and small pixels show their highest relative statistical uncertainty of 1.0$\%$ and 5.5$\%$, respectively. Overall, all data points are within limits. The data points below 10$\,$ms correspond to a potential use case in IR/IC. No issues with the Dosepix are expected for these pulse durations.

\subsection{Dependence on the dose rate}
\begin{figure}[!t]
\centering
\includegraphics[width=3.5in]{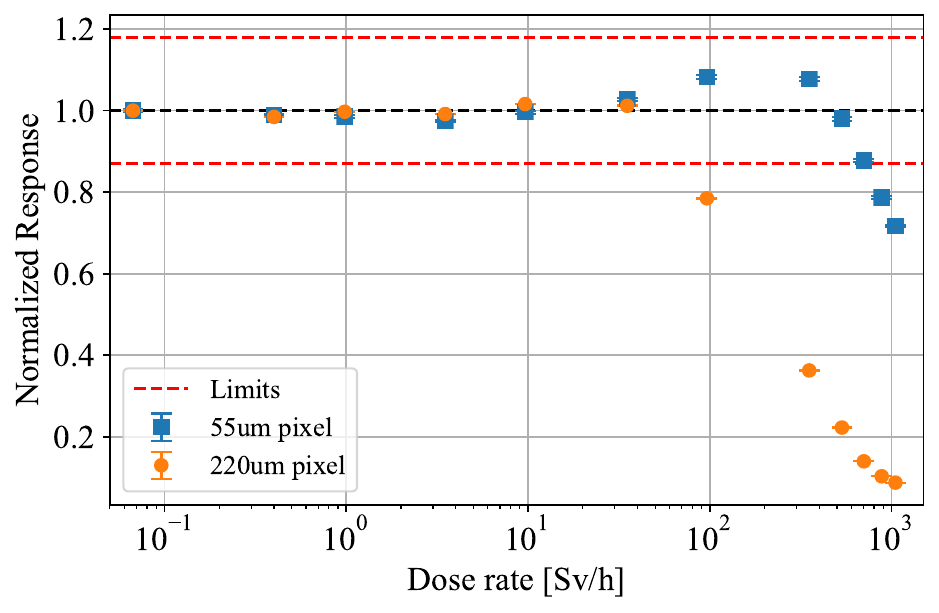}
\caption{Normalized response for $H\mathrm{_p(10)}$ in an RQR8 photon reference field. Influence quantity is the dose rate. The uncertainty bars are calculated via (4).}
\label{pDR}
\end{figure}

The Dosepix dosimetry system was irradiated in the direct radiation field for dose rates in the range of 0.07$\,$Sv/h to 1080$\,$Sv/h. To achieve these dose rates, both the reference dose and the pulse duration were varied. The response for both the small and large pixels was evaluated and is shown in \mbox{Figure \ref{pDR}}. Both pixel sizes have a nearly flat response up to about 35$\,$Sv/h. The normalized response of the large pixels falls below the lower limit at slightly less than 100$\,$Sv/h, whereas for the small pixels, a dose rate up to about 704$\,$Sv/h is achievable in the used reference field. The explanation for the decreasing normalized response with increasing dose rate is pile-up and its corresponding decrease of the count rate. This results in a decreased dose and subsequently a decreased response. The small pixels show a small increase of the normalized response prior to the decrease. The small pixels are less prone to pile up due to their smaller detection volume. An overestimation is observed in the case of several low-energy photons being registered simultaneously, which yields an event with a higher energy. Such an event is sorted into an energy bin a higher conversion factor. Such an overestimation is not observed in this radiation field for the large pixels due a stronger influence of pile up, which results in an overflow of the last bin, and sets an overflow flag. If the flag is set, the bin is not utilized in the dose calculation because it would lead to a significant overestimation of the dose. The normalized response is constant for the dose rates of scattered radiation fields in IR/IC. The spectrum of the scattered radiation field is softer than the spectrum of the primary radiation field. The effectively smaller energies do not pose a problem to the response of Dosepix, which operates above 12$\,$keV. The small pixels would allow an active warning in accident situations, e.g., if the person is exposed to the direct X-ray radiation field of a medical diagnostic X-ray tube where dose rates up to 360$\,$Sv/h can occur \cite{Clair}. The maximum relative statistical uncertainties of the large and small pixels are 0.24$\%$ and 0.49$\%$, respectively. The applied reference dose values are 1$\,$mSv and higher, and the corresponding uncertainty limits between 5-6$\%$ \cite{PTBc}. The observed uncertainties stay within limits.

\subsection{Count rate linearity}
\begin{figure}[hptb]
\centering
\includegraphics[width=3.5in]{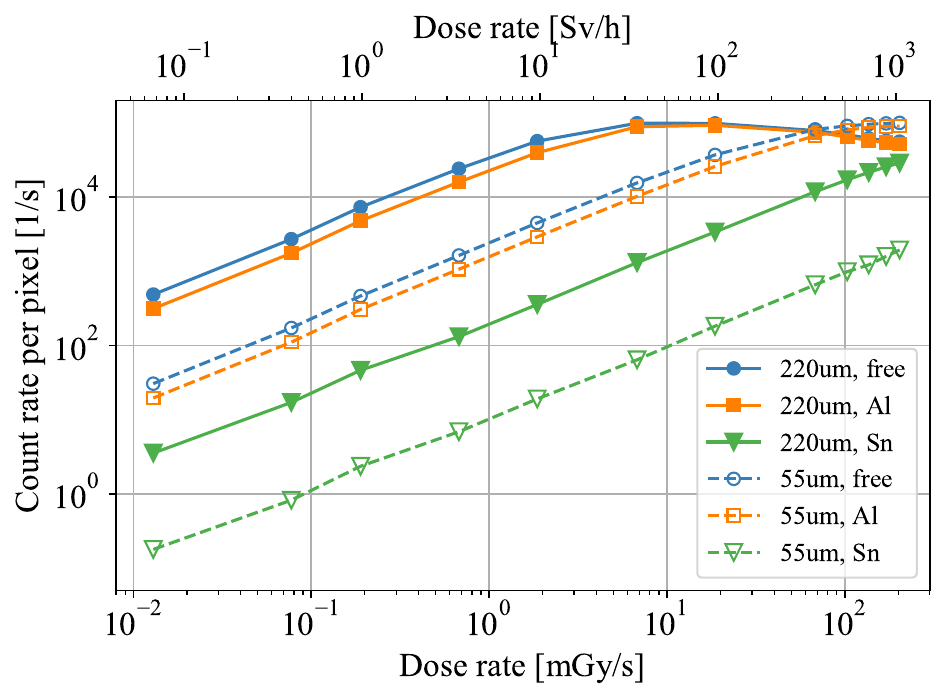}
\caption{Count rate linearity in an RQR8 photon reference field. Influence quantity is the dose rate. The individual count rates of the three differently filtered detectors and their different pixel sizes are shown.}
\label{cDR}
\end{figure}

The measurements with the RQR8 spectrum when varying the dose rate are compared to previous measurements performed in \cite{Zang}. For comparability, the abscissa is first re-scaled to Sv/s and then divided by the $H\mathrm{_p(10)}$/$K_{\mathrm{air}}$ conversion coefficient for the RQR8 spectrum (1.438$\,$Sv/Gy). The results are shown in Figure \ref{cDR} for each pixel size of the three detectors and are additionally labeled by their filter cap, i.e., aluminum cylinder with a thin aluminum foil on its top, which has a hole above the sensor (free), aluminum half-sphere (Al), and tin half-sphere (Sn). Similar behavior is observed as presented by Zang et al.\cite{Zang}, which means that the count rate saturates with high dose rates for unfiltered or weakly filtered detectors. The Dosepix filtered with tin shows for both pixel types no saturation and overall a low count rate. The explanation for the saturation is stated by Zang et al. in arguing that analog pile-up is increasing with an increasing dose rate which is equal to an increase of the flux. Therefore, an increase in the dose rate flattens the deposition spectrum in the detector, namely by converting several low-energy photons into a single high-energy event. For dosimetry, this implies that the dose determination is impacted. A correlation between the normalized response and count rate is observed. From the turning point at 35$\,$Sv/h onward, the response falls below the limit. The compensation of the larger values of the conversion factor in higher energy bins is not strong enough to counteract the loss of separate events. Even due to the count rate saturation and analog pile-up, a good normalized response is achieved. The reason for this is that one of the three detectors - namely the Dosepix filtered with the tin cap - is only in the beginning stage of its saturation. The latter statement implies that its 16 energy bins appropriately represent the energy deposition spectrum and that its partial dose is still correctly determined.

\section{Conclusion}
The Dosepix detector's dependence of the normalized response in an RQR8 pulsed photon field (with a mean energy of 51$\,$keV) was shown for the variation of the pulse duration and the variation of the dose rate. Both tests show promising results for applying the Dosepix detector as an active personal dosemeter measuring the personal dose equivalent $H\mathrm{_p(10)}$ in pulsed photon fields. The pulse duration independence of the dose measured by the prototype of a Dosepix dosimetry system is a direct result of the dead-time-free readout principle of the Dosepix acting as a camera-like radiation detector. Even the shortest pulse durations will not pose any problems to the Dosepix detector, provided that the dose rate during the pulse does not exceed certain limits. As demonstrated here, these limits concerning dose rate are substantial - i.e., in the order of 100$\,$Sv/h and higher - compared to other commercial electronic dosemeters that saturate in the region of a few Sievert per hour. Therefore, it is concluded that the Dosepix detector is a viable detector for dosimetry of pulsed photon fields. The pulsed fields occurring in IR/IC with their parameter space of several milliseconds of pulse duration and a dose rate of up to 10$\,$Sv/h above the lead apron of the operator do not pose any problem to Dosepix. Even for dose rates in the direct beam of the radiation field, a correct indication of the response within limits is expected up to at least 35$\,$Sv/h, and approximately to 70$\,$Sv/h with the large pixels when interpolating between the data points. The small pixels allow a correct indication of the dose up to a factor 10 of the dose rate obtained from large pixels. The large pixels are intended for use in dosimetry, while the small pixels act as a control mechanism. If the dose deviation between large and small pixels exceeds a threshold value, it can be assumed that the dose rate is too high, which would switch the dose measurement to the small pixels until the deviations are below the threshold. The large pixels allow a correct indication of the dose of at least one order of magnitude higher and the small pixels at least two orders of magnitude higher than dosemeters tested in \cite{10.1093/rpd/ncz253,Hupe,10.1093/rpd/ncz173}. It has to be pointed out, that the demonstrated large dose rate range is guaranteed only for the used radiation quality RQR8. In the relevant X-ray energy regime, the detection efficiency is larger for lower energies. On the other hand, the signal pulse length decreases towards lower energies and so does the pile-up probability. The resulting energy dependence of the dose rate limit of our system is therefore not trivial and depends on the spectral shape. Further tests in different energy ranges need to be performed to identify the largest possible dose rate within the legal limits for the normalized response. An application in radiotherapy, e.g., in the scattered radiation field of medical Linac, where photon energies of several hundreds of kilo electron volts up to mega electron volts are expected needs to be investigated separately. If the dose rates in the scattered beam are within the ranges investigated in this publication, then an application of Dosepix in Linac is feasible. Comparative studies to already existing pixel detectors which allow energy resolved dosimetry as already investigated in \cite{microdosimetry} must also be carried out in the future.

\bibliographystyle{IEEEtran}
\bibliography{IEEEabrv,main.bib}
\end{document}